# PQC standards alternatives: reliable semantically secure key encapsulation mechanism and digital signature protocols using the rank-deficient matrix power function

**Juan Pedro Hecht** [1], **Hugo Daniel Scolnik** [1,2,*]

[1.] Master's degree in Information Security, Economic Science School, School of Exact and Natural Sciences and Engineering School, University of Buenos Aires, phecht@dc.uba.ar

ORCID 0000-0002-4196-9059

[2] [*]Computer Science Department, School of Exact and Natural Sciences, University of Buenos Aires, hugo@dc.uba.ar, [*]corresponding author.

ORCID 0000-0002-0684-3661

**Abstract:** post-quantum cryptography (PQC) aims to develop public-key primitives that are secure against adversaries using classical and quantum computing technologies. This study introduces novel protocols, a key encapsulation mechanism, a digital signature scheme, and special protection against linear attacks. Our purpose is to create reliable alternatives to current standards, seeking compact, fast, and secure replacements of the key interchange and digital signature in the TLS 1.3 protocol, which safeguards Internet traffic, allowing an easy post-quantum transition to protect current data from the "*harvest now, decrypt later*" threat.

**Keywords** KEM, DSA, non-commutative algebraic cryptography, PQC, post-quantum cryptography, provable cryptography, semantic security, IND-CCA2, UF-CMA, rank-deficient matrices, matrix power function.

# Introduction

Post-quantum cryptography (PQC) [1] seeks efficient public-key primitives that remain secure against adversaries using quantum computers. Among the most prominent approaches are lattice-based, code-based, isogeny-based, multivariate quadratic-based and algebraic matrix-based frameworks. The Matrix Power Function [2, 3–5] paradigm has emerged within the latter category as a versatile algebraic building block for key establishment and authentication. However, it is susceptible to linear attacks. In [2], we presented the Rank-Deficient Matrix Power Function (RDMPF), which is modified using a sigma variant and rank-deficient matrices to prevent these types of attack.

The security of our protocols relies on the Random Oracle Model (ROM), the current standard. We achieve this by using the Fujisaki-Okamoto transform (FO) [6] and implicit rejection (IR)[7] techniques, which provide a provable level of security for our proposals when used together. This study focuses on two cryptographic protocols derived from the RDMPF framework. The first is FO-RDMPF-KEM, a key encapsulation mechanism (KEM) constructed from the RDMPF primitive. It is enhanced via ciphertexts or signatures that yield pseudorandom outputs, mitigating side-channel and random oracle-based leakage. The second protocol, FO-RDMPF-DSA, is a digital signature algorithm (DSA) based on the same framework and principles. Together, these two protocols constitute the minimum set required for post-quantum cryptography (PQC) securitization of Transport Layer Security (TLS) 1.3, the cryptographic keystone of internet traffic.

The KeyGen() one-way trapdoor function (OWTF) is described in our paper [2] as the second protocol, which runs over R rounds. The correct definition of the (sk, pk) data can be found in the Mathematica code for the toy example. The RDMPF is outlined in the next paragraph to improve the readability of this work.

# RDMPF explanation

This is the core matrix function used in the proposed protocols. Here, "RD" stands for "rank-deficient" matrices, which are a special class of matrices that help prevent classical linearization attacks [2]. MPF stands for "matrix power function," which is an exponential-like extension of standard matrix multiplication. No classical or quantum attacks have yet been reported against the generated outputs, making this approach a strong candidate for secure implementation. The overall structure is shown in Table 1.

**RDMPF function (sigma variation)**

```
RDMPF[X_,W_,Y_]:= Module[{i, j, L, K, z, ex, pr}, (* local variables *)
Q=Rmat[rows, cols, prime]; (* modular random integer matrix, here rows=cols=dim *)
Do[
  Do[
    pr=1;
    Do[
      Do[
        ex=Mod[Times[sigma, Times[X[[i,K]],Y[[L,j]]]], prime-1]; (*modular product *)
        z=IntFastPower[W[[K,L]], ex]; (* square-and-multiply power function*)
        pr=Times[pr, z];
      {L, 1, cols}]; (* end Do *)
    {K, 1, cols}]; (* end Do *)
    Q[[i, j]]=Mod[pr, prime];
  {j, cols}] (* end Do *)
{i, rows}]; (* end Do *)
Q]; (* output *)
```

**Table 1.** shows the pseudocode of the RDMPF function. The output is a Q matrix that is too complex to write in closed form. Only the Qi,j components could be presented [2].

To take a feeling of the complexity involved, we present the following symbolic example:

```
rows = cols = 3;
W = {{w11, w12, w13}, {w21, w22, w23}, {w31, w32, w33}};
X = {{x11, x12, x13}, {x21, x22, x23}, {x31, x32, x33}};
Y = {{y11, y12, y13}, {y21, y22, y23}, {y31, y32, y33}};
Q = {{q11, q12, q13}, {q21, q22, q23}, {q31, q32, q33}};
Q = RDMPF[X, W, Y]; (* using sigmafunction *)
Print[Q]; (* each line is a Q matrix term *);
```

$$
\begin{aligned}
\{\{ & w_{11}^{\sigma x_{11} y_{11}} w_{12}^{\sigma x_{11} y_{21}} w_{13}^{\sigma x_{11} y_{31}} w_{21}^{\sigma x_{12} y_{11}} w_{22}^{\sigma x_{12} y_{21}} w_{23}^{\sigma x_{12} y_{31}} w_{31}^{\sigma x_{13} y_{11}} w_{32}^{\sigma x_{13} y_{21}} w_{33}^{\sigma x_{13} y_{31}}, \\
& w_{11}^{\sigma x_{11} y_{12}} w_{12}^{\sigma x_{11} y_{22}} w_{13}^{\sigma x_{11} y_{32}} w_{21}^{\sigma x_{12} y_{12}} w_{22}^{\sigma x_{12} y_{22}} w_{23}^{\sigma x_{12} y_{32}} w_{31}^{\sigma x_{13} y_{12}} w_{32}^{\sigma x_{13} y_{22}} w_{33}^{\sigma x_{13} y_{32}}, \\
& w_{11}^{\sigma x_{11} y_{13}} w_{12}^{\sigma x_{11} y_{23}} w_{13}^{\sigma x_{11} y_{33}} w_{21}^{\sigma x_{12} y_{13}} w_{22}^{\sigma x_{12} y_{23}} w_{23}^{\sigma x_{12} y_{33}} w_{31}^{\sigma x_{13} y_{13}} w_{32}^{\sigma x_{13} y_{23}} w_{33}^{\sigma x_{13} y_{33}}\}, \\
\{ & w_{11}^{\sigma x_{21} y_{11}} w_{12}^{\sigma x_{21} y_{21}} w_{13}^{\sigma x_{21} y_{31}} w_{21}^{\sigma x_{22} y_{11}} w_{22}^{\sigma x_{22} y_{21}} w_{23}^{\sigma x_{22} y_{31}} w_{31}^{\sigma x_{23} y_{11}} w_{32}^{\sigma x_{23} y_{21}} w_{33}^{\sigma x_{23} y_{31}}, \\
& w_{11}^{\sigma x_{21} y_{12}} w_{12}^{\sigma x_{21} y_{22}} w_{13}^{\sigma x_{21} y_{32}} w_{21}^{\sigma x_{22} y_{12}} w_{22}^{\sigma x_{22} y_{22}} w_{23}^{\sigma x_{22} y_{32}} w_{31}^{\sigma x_{23} y_{12}} w_{32}^{\sigma x_{23} y_{22}} w_{33}^{\sigma x_{23} y_{32}}, \\
& w_{11}^{\sigma x_{21} y_{13}} w_{12}^{\sigma x_{21} y_{23}} w_{13}^{\sigma x_{21} y_{33}} w_{21}^{\sigma x_{22} y_{13}} w_{22}^{\sigma x_{22} y_{23}} w_{23}^{\sigma x_{22} y_{33}} w_{31}^{\sigma x_{23} y_{13}} w_{32}^{\sigma x_{23} y_{23}} w_{33}^{\sigma x_{23} y_{33}}\}, \\
\{ & w_{11}^{\sigma x_{31} y_{11}} w_{12}^{\sigma x_{31} y_{21}} w_{13}^{\sigma x_{31} y_{31}} w_{21}^{\sigma x_{32} y_{11}} w_{22}^{\sigma x_{32} y_{21}} w_{23}^{\sigma x_{32} y_{31}} w_{31}^{\sigma x_{33} y_{11}} w_{32}^{\sigma x_{33} y_{21}} w_{33}^{\sigma x_{33} y_{31}}, \\
& w_{11}^{\sigma x_{31} y_{12}} w_{12}^{\sigma x_{31} y_{22}} w_{13}^{\sigma x_{31} y_{32}} w_{21}^{\sigma x_{32} y_{12}} w_{22}^{\sigma x_{32} y_{22}} w_{23}^{\sigma x_{32} y_{32}} w_{31}^{\sigma x_{33} y_{12}} w_{32}^{\sigma x_{33} y_{22}} w_{33}^{\sigma x_{33} y_{32}}, \\
& w_{11}^{\sigma x_{31} y_{13}} w_{12}^{\sigma x_{31} y_{23}} w_{13}^{\sigma x_{31} y_{33}} w_{21}^{\sigma x_{32} y_{13}} w_{22}^{\sigma x_{32} y_{23}} w_{23}^{\sigma x_{32} y_{33}} w_{31}^{\sigma x_{33} y_{13}} w_{32}^{\sigma x_{33} y_{23}} w_{33}^{\sigma x_{33} y_{33}}\}\}
\end{aligned}
$$

**Figure 1.** shows the RDMPF(X, W, Y) function in symbolic form. Using three matrices highlights the structural complexity of the resulting matrix, which is not feasible in numeric form. It is not feasible to express Q in a simple closed form; however, each of its nine components can be represented as a combination of basic operations. For example, $Q_{1,1}$ and $Q_{3,3}$ correspond to the first and last rows in the printed output.

# Results

First, we provide the formal definitions of the KEM and DSA algorithms in accordance with the restrictions outlined in our previous study [2]. These definitions follow the style conventions used in FIPS 203, i.e., stepwise-structured pseudocode. Since this type of authenticated key agreement relies on asymmetric cryptography, the initial step involves developing a secret key (SK) and its corresponding public key (PK) using the RDMPF post-quantum function.

**KEY ENCAPSULATION MECHANISM**

---

**Algorithm 1: RDMPF-KEM**
***PQC Rank-Deficient Matrix Power Function – second protocol KEM) with FO transform and implicit rejection .***

---

**Key Generation**

1. (pk, sk) ← KeyGen()
2. return (pk, sk).

**•Encapsulation**

3. m ←$ {0,1}^κ
4. (X_r, Y_r)_{r=1..R} ← MapToXY(m)
5. For r = 1..R do
6. TA_r ← RDMPF(X_r, W, Y_r)
7. S_r ← RDMPF(X_r, TB_r, Y_r)
8. end for
9. TA_enc ← Encode(TA_1,...,TA_R)
10. S_enc ← Encode(S_1,...,S_R)
11. Z ← KDF(S_enc)
12. mask ← $H_1$(Z || TA_enc || pk)
13. c_mask ← m XOR mask[0..κ-1]
14. tag ← $H_2$(m || TA_enc || pk)
15. ct := (TA_enc, c_mask, tag)
16. K ← KDF(Z || ct || 0x00)
17. Output (ct, K)

**Decapsulation**

1. Parse ct into (TA_enc, c_mask, tag).
2. If sk is a seed, then expand → (U_r, V_r)_{r=1..R}.
3. For r = 1..R do
4. S'_r ← RDMPF(U_r, TA_r, V_r)
5. end for
6. S'_enc ← Encode(S'_1,...,S'_R)
7. Z' ← KDF(S'_enc)
8. mask' ← $H_1$(Z' || TA_enc || pk)
9. m' ← c_mask XOR mask'[0..κ-1]
10. (X'_r, Y'_r) ← MapToXY(m')
11. For r = 1..R do
12. TA"_r ← RDMPF(X'_r, W, Y'_r)
13. end for
14. TA"_enc ← Encode(TA"_1,...,TA"_R)
15. tag' ← $H_2$(m' || TA_enc || pk)
16. If (TA"_enc = TA_enc) ∧ (tag' = tag) then
17. K ← KDF(Z' || ct || 0x00)
18. return K
19. Otherwise
20. ρ ← $H_1$(SecretFallback(sk) || ct || 0xFF)
21. K ← KDF(ρ || ct || 0x01)
22. return K

---

**Correctness**
For all (pk, sk) generated by KeyGen() and (ct, K) generated by Encaps(K, pk), the following is true: Decaps(sk, ct) = K.

**Proposition (Security of RDMPF-KEM)**
Assume that the underlying RDMPF-PKE primitive is IND-CPA-secure. Then, in the Random Oracle Model (ROM), the scheme Π_(RDMPF-KEM) is IND-CCA2-secure [10].

**Proof Sketch.**
The proof sketch covers the key concepts: $H_1$ for derandomization/re-encryption (FO), and the failure key/implicit rejection to prevent oracle leakage.

$H_1$ derandomizes the encapsulation, turning it into a deterministic function of m and pk. This enables the decapsulator to perform re-encryption verification. Using a failure key guarantees that invalid ciphertexts generate pseudorandom outputs, thus eliminating any useful decryption oracle function. Including ct in the derivation of K prevents substitution attacks. It is crucial to ensure that the value returned to the external verifier is always a fixed-length string that is indistinguishable from a valid tag. The verifier receives a pseudorandom value that fails the internal check and requires constant time to compute.

**DIGITAL SIGNATURE ALGORITHM (DSA)**
The second algorithm illustrates the DSA procedure. It describes the Fujisaki–Okamoto transform (FO-DSA) [8] for digital signatures with implicit rejection [9]. It follows the stepwise style conventions used in FIPS 204 [11].

**Setup and primitives**
Let Π_{SIG} = (KeyGen, Sign, Verify) be a randomized digital signature scheme where Sign(sk, M; r) uses explicit randomness r, and Assume that Sign is UF-CMA secure.

**Let:**
$H_1$: {0,1}* → {0,1}^ρ is a hash function (random oracle) used to derive the signing randomness.
$H_2$: {0,1}* → {0,1}^κ is a hash function used to derive authentication tags.

A fixed secret value, z, chosen once during initialization, is used in rejection cases. Here, ρ denotes the random length of the original signature algorithm, κ denotes the tag length, and ⊥ denotes invalid output. The concatenations are bit strings.

---

**Algorithm 2: FO-DSA = (KeyGen, Sign, Verify)**
***PQC Rank-Deficient Matrix Power Function – second protocol KEM) with FO transform and implicit rejection.***

---

**Key Generation**
1. KeyGen_{DS}():
2. (pk, sk) ← KeyGen()
3. return (pk, sk)

**Signing (Deterministic)**
1. Sign_{DS}(sk, M):
2. r ← $H_1$('r' || M || pk)
3. $\sigma_0$ ← Sign(sk, M; r)
4. t ← $H_2$('t' || $\sigma_0$ || M || pk)
5. return σ := ($\sigma_0$, t)

**Verification with Implicit Rejection**

1. Verify_{DS}(pk, M, σ):
2. Parse σ as ($\sigma_0$, t)
3. valid ← Verify(pk, M, $\sigma_0$)
4. t' ← $H_2$('t' || $\sigma_0$ || M || pk)
5. If valid = accept and t' = t then
6. return accept
7. else return reject* // [(a)] pseudorandom placeholder = $H_2$('z' || z || $\sigma_0$ || M)

---

**Correctness**

For all (pk, sk) generated by KeyGen() and all messages M, if σ is generated by Sign_(DS)(sk, M), then Verify_(DS)(pk, M, σ) returns "accept"; otherwise, it returns "reject."

**Proposition (Security of FO-DSA with Implicit Rejection)**

Assuming P_(SIG) is UF-CMA secure and $H_1$ and $H_2$ behave as random oracles [12], the transformed scheme P_(FO-DS-IR) is UF-CMA secure. Implicit rejection ensures that even malformed signatures yield pseudorandom outputs, thus preventing information leakage.

**Proof Sketch**.

The deterministic randomness of $H_1$ prevents nonce misuse. Tag t from $H_2$ binds the signature to the message m. Implicit rejection maps invalid signatures to pseudorandom outputs via z, which eliminates any adversarial advantages. Any forgery would require breaking either the signature scheme (Π_(SIG)) or the random oracle model.

**Implementation Notes**

- Maintain strict domain separation ("r", "t", "z")

- Use constant-time verification to avoid side-channel leaks.

[(a)] Reject* must be a fixed-format pseudorandom value that matches the correct answer, not just an error value. Ensure that the value returned to the external verifier is always a fixed-length string indistinguishable from a valid tag.

**SECURITY OF THE RDMPF**

The complexity of the MPF core function used in the KEM and DSA protocols (see Tables 1 and 2, and Figures 1 and 2) renders a quantum attack based on Shor's algorithm infeasible. Therefore, an attacker can only attempt a classical attack, such as a linearization attack [3], which is blocked by singular matrices of deficient rank and the sigma variant described here. The remaining task is to prevent a brute-force attack, i.e., recovering the private key (sk) in KeyGen() based on the public key (pk) or systematically trying the private matrices W, X, and Y.

If the dimension of these matrices is n (for W) or $n - 1$ (for X and Y), $3n^2 - 4n + 2$ unknowns are generated, which is an expression of complexity $O(n^2)$. Taking these factors into account, the security of the KEM and DSA protocols can be estimated as shown in Table 2.

| Matrix dimension (n) | Unknown values | Unknown value bit size | Total bit security | NIST PQC classical security level |
|---|---|---|---|---|
| 3 | 17 | 32 | 544 | 5 (= 512-bit) |
| 5 | 57 | 32 | 1824 | 5 |
| 7 | 121 | 32 | 3872 | 5 |
| 10 | 262 | 32 | 8384 | 5 |
| 15 | 617 | 32 | 19744 | 5 |
| 20 | 1122 | 32 | 35904 | 5 |

**Table 2.** shows the proposed security levels for the core MPF function in the event of quantum and linearization attacks. These results are expected to correspond to brute-force attacks against our protocols. The PQC quantum security level is the square root of the classical value due to the Grover algorithm [1].

For real-life implementation, it is suggested to use a dimension n greater or equal to seven.

## TRANSITION FROM ALGORITHMIC SPECIFICATION TO EXPERIMENTAL VALIDATION

The stepwise pseudocode descriptions of FO-RDMPF-KEM and FO-RDMPF-DSA show that these constructions are structurally correct and secure in the Random Oracle Model. To substantiate these constructions beyond formal analysis, we developed reference implementations in Mathematica 12.1+ and conducted controlled runtime experiments on an x86_64 architecture.

These implementations adhere directly to the algorithmic definitions, and the corresponding results are presented in Appendix A.

· **Key generation** instantiates the Rank-Deficient Matrix Power Function (RDMPF) within the one-way trapdoor function framework to derive (pk, sk) pairs.

· **Encapsulation/decapsulation** uses the FO transform and implicit rejection to ensure deterministic re-encryption consistency and ciphertext indistinguishability under a chosen ciphertext attack.

· **Signing and verification** incorporate deterministic randomness via hash-derived nonces to prevent misuse of randomness, while implicit rejection enforces pseudorandom responses to malformed inputs.

Although the toy implementations are realized in an interpreted, unoptimized environment, they preserve the logical structure of the protocols. This allows us to verify correctness properties such as key agreement consistency, signature integrity, and tamper detection through implicit rejection while providing performance estimates of the constituent operations (KeyGen, Encaps, Decaps, Sign, and Verify).

Subsequent experimental results confirmed the functional soundness of our protocols for explicit test cases. These results demonstrate that FO–RDMPF constructions, although algebraically distinct, are computationally viable and amenable to further optimization in a deployment-grade implementation.

Appendix B shows the results of the preliminary comparisons. A brief analysis of the obtained data is presented below.

**For Kyber (ML-KEM)** on x86_64, wall-clock times are expected to be in the microsecond range (see the wolfSSL Linux numbers above: ~4–14 µs per KeyGen/Enc/Decap, depending on the parameter set and implementation) [13].

**For Dilithium (ML-DSA)** on x86_64, we can expect millisecond-level signing and verification times. Published studies report ≤15 ms for sign/verify in real TLS experiments, but exact numbers vary by parameter set.

The Mathematica results confirm the algebraic feasibility and provide a baseline cost analysis for the matrix operations.

It must be emphasized that times in the millisecond to microsecond range are irrelevant in real-life applications, as these times go unnoticed when keys are interchanged or transactions are signed amidst internet traffic. More attention should be paid to resilience against such attacks. Efforts should be directed toward developing resilient PQC protocols with proven security levels.

# Discussion

The algebraic simplicity of RDMPF, compared to the polynomial arithmetic of module lattices (NIST's solutions), suggests strong performance potential. However, this potential can only be realized through future optimized C/Rust/assembly implementations

FO-RDMPF constructions combine algebraic transparency with modern cryptographic hardening techniques, such as the Fujisaki-Okamoto transform and implicit rejection, as recommended by NIST. The results confirm several key aspects of this research.

**Security Amplification via FO Transformation and IR rejection.**
This elevates the RDMPF primitives to a stronger level, ensuring resilience against adaptive chosen-ciphertext and forgery attacks in the random oracle model. The system achieves an IND-CCA2 security level for KEM and a UF-CMA level for DSA

A key feature of both protocols is replacing error flags with pseudorandom outputs. This eliminates oracle leakage and simplifies implementation. Our toy experiments confirmed that this can be realized efficiently, as can the UF design.

**Algebraic Efficiency**
The RDMPF framework relies on rank-deficient matrix power computations, which are structurally simpler than lattice-based reductions. Even in symbolic Mathematica demonstrations, these operations are fast and scalable. With optimized implementations, such as modular linear algebra over GF(p^m), the schemes are expected to perform as well as state-of-the-art PQC candidates.

**Reliability and Practical Implications**
By eliminating nonce requirements (i.e., derandomized encapsulation and signing), the protocols mitigate the risks associated with randomness failures. Historically, randomness failures have been a critical vulnerability in signature schemes. This enhances the robustness of constrained environments.

**Transition to Post-Quantum Cryptography Standards**
NIST published an internal report (NISTIR8547) [14] analyzing fast hybrid solutions to attain a future PQC universal framework. The transition to PQC involves integrating quantum-resistant algorithms into TLS 1.3. This flexible foundation enables the seamless integration of new algorithms and supports perfect forward secrecy (PFS) to counter the "*harvest now, decrypt later*" threat [15]. This process requires updating TLS 1.3 cryptographic libraries to PQC-ready stacks and implementing hybrid approaches that combine classical and PQC algorithms (see references [16, 17]). It may also entail adopting protocols such as RDMPF-KEM and RDMPF-DSA.

# Future Directions

Although the toy examples demonstrate the correctness and feasibility of the approach, more work is needed to increase the size of the parameters to reach post-quantum security levels. Side-channel and fault-injection resistance analyses must be conducted. In summary, FO-RDMPF-KEM and FO-RDMPF-DSA are reasonable, fast, reliable, and semantically secure PQC protocols. They are grounded in algebraic simplicity but strengthened by modern cryptographic transformations. Symbolic implementations confirm their soundness and pave the way for deployment-oriented evaluations. These protocols can efficiently replace ML-KEM (Crystals Kyber) and ML-DSA (Dilithium) [18, 19].

In conclusion, rigorous side-channel and fault-injection resistance analyses should be conducted; the parameter sets should be scaled to align with NIST security levels; and deployment-grade implementations should be developed in C or Rust. These steps are essential for obtaining a fair performance comparison with the optimized lattice standards.

Author contributions**:** both authors contributed equally to this study.

The authors declare no conflicts of interest.

# References

[1] Bernstein D. et al. (2009), Post-Quantum Cryptography, Springer, https://doi.org/10.1007/978-3-540-88702-7
[2] Hecht J. P., Scolnik, H. D. (2025), Post-Quantum Key Agreement Protocols Based on Modified Matrix-Power Functions over Singular Random Integer Matrix Semirings, Computer Networks and Communications, 3:1, 1-18
[3] Sakalauskas E., Mihalkovich A., (2017), Improved Asymmetric Cipher Based on Matrix Power Function Resistant to Linear Algebra Attack, Vilnius University, 28:3, DOI: http://dx.doi.org/10.15388/Informatica.2017.142
[4] Sakalauskas E. and Mihalkovich A. (2018), MPF Problem over Modified Medial Semigroup Is NP-Complete. Symmetry 10, 571.
[5] Sakalauskas E. and Mihalkovich A. (2020), A New asymmetric cipher of non-commuting cryptography class based on enhanced MPF, IET Information Security, Volume 14, Issue 4, 410-418
[6] Fujisaki E, Okamoto T (2013) Secure integration of asymmetric and symmetric encryption schemes. Journal of Cryptology 26:80–101. https://doi.org/10.1007/s00145-011-9114-1
[7] Joux , A., Loss, J., & Wagner, B. (2025, May). Kleptographic attacks on implicit rejection. In IACR International Conference on Public-Key Cryptography (pp. 214-245). Cham: Springer Nature Switzerland.
[8] NIST FIPS 203 (2024) - Module-Lattice-Based Key-Encapsulation Mechanism Standard, https://nvlpubs.nist.gov/nistpubs/FIPS/NIST.FIPS.203.pdf
[9] Menezes, A. J., van Oorschot, P. C., & Vanstone, S. A. (1996). Handbook of applied cryptography, CRC Press
[10] Bogdanov, D. (2005). IND-CCA2 secure cryptosystems. University of Tartu.
[11] NIST FIPS 204 (2024) - Module- Lattice - Based Digital Signature Standard, https://nvlpubs.nist.gov/nistpubs/FIPS/NIST.FIPS.204.pdf
[12] Koblitz, N., & Menezes, A. J. (2015). The random oracle model: a twenty-year retrospective study Designs, Codes and Cryptography , 77 (2), 587-610.
[13] WolfSSL Linux benchmark ML-KEM / Kyber (x86_64 ) Intel i7- 1185G7 @ ~3.00 GHz, single core, https://www.wolfssl.com/post-quantum-kyber-benchmarks-linux/ (consulted 2025 9.,11)

[14] NIST Internal Report, NIST IR 8547 ipd (2024) , Transition to Post-Quantum Cryptography, Standards, https://nvlpubs.nist.gov/nistpubs/ir/2024/NIST.IR.8547.ipd.pdf

[15] Olutimehin , A. T., Joseph, S., Ajayi, A. J., Metibemu , O. C., Balogun, A. Y., & Olaniyi, O. O. (2025). Future-proofing data: Assessing the feasibility of post-quantum cryptographic algorithms to mitigate 'harvest now, decrypt later' attacks.

[16] Sosnowski, M., Wiedner, F., Hauser, E., Steger, L., Schoinianakis, D., Gallenmüller, S., & Carle, G. (2023, December). Performance of post-quantum TLS 1.3. In Companion of the 19th International Conference on Emerging Networking Experiments and Technologies (pp. 19-27).

[17] NDSS 2020, Post-Quantum Authentication in TLS 1.3: A Performance Study.

[18] Migliore, V., Gérard, B., Tibouchi, M., & Fouque, P. A. (2019), May). Masking dilithium: Efficient implementation and side-channel evaluation. In the Proceedings of the International Conference on Applied Cryptography and Network Security (pp. 344-362). Cham: Springer International Publishing.

[19] Mao, G., Chen, D., Li, G., Dai, W., Sanka, A. I., Koç, Ç. K., & Cheung, R. C. (2023). High-performance and configurable SW/HW co-design of post-quantum signature CRYSTALS-DILITHIUM. ACM Transactions on Reconfigurable Technology and Systems, 16(3), 1-28.

# APPENDIX A

## Brief glossary and simple explanatory concepts

**Digital Signature Algorithm (DSA):** A standard procedure that generates a string of bits that certifies the origin, integrity, and non-repudiation of any document.

**FIPS:** U.S. government cryptographic standards for secure information processing issued by the NIST.

**FO (Fujisaki–Okamoto Transform):** Technique converting public-key encryption into an IND-CCA2–secure scheme through hashing and re-encryption.

**GF(p) (finite prime field, p=prime number):** A closed integer number set over which sums and products are realized.

**HASH:** A relatively small string of bits, usually 256 to 1024 bits in length, that compact any message in such a way that it can detect any change.

**IND-CCA2:** indistinguishable adaptative chosen ciphertext attack, strongest encryption security notion, where ciphertexts remain indistinguishable despite adaptive decryption queries.

**IR (Implicit Rejection):** A decryption method that silently rejects invalid ciphertexts to prevent side-channel leakage.

**Key Derivation Function (KDF):** An algorithm that takes some initial secret (such as a password, passphrase, or shared secret key) and produces one or more strong cryptographic keys.

**KEM (Key-Encapsulation Mechanism):** Primitive for securely transmitting symmetric session keys via public-key encryption.

**ML (Module Lattice):** Algebraic lattice structure over modules underpinning efficient post-quantum schemes (e.g., ML-KEM, ML-DSA).

**NIST (National Institute of Standards and Technology):** A U.S. government agency, part of the Department of Commerce, that develops and promotes measurement standards, technology, and security guidelines.

**NONCE (No More Than Once):** a piece of random information that should be used only once in any session.

**OWTF (One-Way Trapdoor Function):** A special type of apparent irreversible function used as the mathematical core of public-key cryptography.

**Post-Quantum Cryptography (PQC):** Cryptography designed to remain secure against adversaries equipped with future quantum computers.

**Rank-Deficient Matrix Power Function (RDMPF):** Transformation mapping of standard matrix multiplication into an exponential-like form using rank-deficient matrices for increased cryptanalytic resistance.

**Random Oracle Model (ROM):** an idealized model that treats hash functions as deterministic, yet perfectly random oracles for security proofs.

**SS (Semantic Security):** The notion of ensuring that ciphertexts reveal no partial information about plaintexts (equivalent to IND-CPA).

**UF-CMA:** Signature security notion preventing forgeries, even after exposure to chosen message signatures.

**XOF (Extendable-Output Function)**: This is a generalization of a hash function; instead of outputting a fixed length, an XOF can produce an output of any length.

**XOR (Exclusive Or):** a logical operation over bits.

# APPENDIX B

## Experimental KEM and DSA correctness and preliminary runtime benchmarks.

The experimental setup parameters for the elaboration of the tables below are as follows:

**Architecture:** X86_64 bits
**Computer:** Intel Core i5-5200U CPU @2.20 GHz 256 GB SSD RAM 8GB ROM + Win 10
**Program environment**: Mathematica 12.1.0.0., interpreted unoptimized code
**Algoritm:** RDMPF KEM and RDMPF-DSA with Fujisaki-Okamoto transform and Implicit Rejection
**Semantic security level**: IND-CCA2 (KEM) and UF-CMA (DSA)
**KEM Parameters**: R (rounds)=1, (matrix) dim=5, GF(997) prime field, expMax (powers)=9, k = 64, sigma=3
**DSA Parameters**: R (rounds)=1, (matrix) dim=5, GF(997) prime field, k = 64, sigma=3

| RUN # | KEYa = KEYb | Setup() time \ (seconds) | a.K eyGen() time (seconds) | b.Encaps() time (seconds) | c.Decaps() time (seconds) | c*.ImplicitReject() time (seconds) | a.b.c. time (seconds) |
|---|---|---|---|---|---|---|---|
| 1 | Yes | 0.0156259 | 0.0624936 | 0.0781216 | 0.0312470 | 0.0312469 | 0.1562371 |
| 2 | Yes | 0.0312331 | 0.0937425 | 0.0695100 | 0.0468730 | 0.0156263 | 0.2101255 |
| 3 | Yes | 0.0156251 | 0.0624958 | 0.0781191 | 0.0312475 | 0.0312485 | 0.1718624 |
| 4 | Yes | 0.0312459 | 0.1093696 | 0.0781181 | 0.0468716 | 0.0156232 | 0.2343593 |
| 5 | Yes | 0.0156186 | 0.0624984 | 0.0624927 | 0.0312491 | 0.0156219 | 0.1562402 |
| 6 | Yes | 0.0624957 | 0.2343596 | 0.1249909 | 0.0312491 | 0.0312478 | 0.3905996 |
| 7 | Yes | 0.0312484 | 0.0624973 | 0.0937418 | 0.0355995 | 0.0156253 | 0.1918386 |
| 8 | Yes | 0.0212867 | 0.0624946 | 0.0624979 | 0.0312455 | 0.0156275 | 0.1562380 |
| 9 | Yes | 0.0156230 | 0.0624969 | 0.0624970 | 0.0156256 | 0.0156218 | 0.1406195 |
| 10 | Yes | 0.0312319 | 0.0468674 | 0.0937424 | 0.0312469 | 0.0312531 | 0.1718567 |
| | **Mean** | 0.0271234 | 0.0859316 | 0.0803832 | 0.0332455 | 0.0218742 | 0.1979977 |
| | **Standard error** | 0.0043383 | 0.0165718 | 0.0058654 | 0.0026671 | 0.0024206 | 0.0220211 |

**Table 1:** Correctness and running times of Algorithm 1: RDMPF-KEM Toy Demo. In real-life applications, the implicit rejection time (c*) should be equal to the decaps time (c) to prevent side-channel leakage.

**KEM TOY DEMO FINAL SUMMARY OUTPUT**

The source code written in Mathematica 12 and full output can be downloaded from https://1drv.ms/f/c/32525337ca8f82b9/EgW4CnRWjUhLuDrokCd5rkABQwE83kSpQC-teRb2-OD8qQ?e=aWIehz

Session Key (encaps): 1100111110010010
Session Key (decaps): 1100111110010010
Session Keys Match: True
Tampering Test: PASSED
Protocol Status: SUCCESS

=== **SECURITY FEATURES HIGHLIGHTED** ===

FUJISAKI-OKAMOTO TRANSFORM: Ensured by deterministic hashing

. Transforms the OW-CPA scheme into IND-CCA2 KEM

. Use of hash functions for consistent key derivation

. Integrity verification using tags

IMPLICIT REJECTION: Fully implemented

. Rejection of invalid ciphertext: incorrect format

. Rejection of tag verification failed: manipulation detected

. Rejection key generation using Secret Fallback

. Attacker cannot distinguish the rejection cause

=== **KEM DEMO COMPLETE** ===

Elapsed computing time (seconds) = 0.2187312

| RUN # | Signature Verification Result | Setup() time (seconds) | a.Signature() time (seconds) | b.Verification() time (seconds) | c*.ImplicitReject() time (seconds) | a.b. time (seconds) |
|---|---|---|---|---|---|---|
| 1 | OK | 0.2031101 | 0.1562408 | 0.0156234 | 0.0156247 | 0.1718642 |
| 2 | OK | 0.2187352 | 0.2031110 | 0.0156228 | 0.0312485 | 0.2187338 |
| 3 | OK | 0.1874876 | 0.2917030 | 0.0624966 | 0.0156237 | 0.3541996 |
| 4 | OK | 0.2968546 | 0.1093704 | 0.0312444 | 0.0156244 | 0.1406148 |
| 5 | OK | 0.1249895 | 0.0981242 | 0.0156206 | 0.0156244 | 0.1137448 |
| 6 | OK | 0.1249919 | 0.0937460 | 0.0156207 | 0.0156258 | 0.1093667 |
| 7 | OK | 0.1718600 | 0.1249917 | 0.0156252 | 0.0156225 | 0.1406169 |
| 8 | OK | 0.0937426 | 0.0624987 | 0.0156213 | 0.0156256 | 0.0781200 |
| 9 | OK | 0.1093620 | 0.0624965 | 0.0156266 | 0.0156207 | 0.0781231 |
| 10 | OK | 0.0781226 | 0.0624934 | 0.0156297 | 0.0156297 | 0.0781231 |
| | **Mean** | **0.1609256** | **0.1264776** | **0.0218731** | **0.0171870** | **0.1483507** |
| | **Standard error** | **0.0202570** | **0.0220247** | **0.0045283** | **0.0014822** | **0.0255956** |

**Table 2:** Correctness and running times of Algorithm 2: RDMPF-DSA Toy Demo. In real-life applications, the implicit rejection time (c*) should be equal to the verification time (b) to prevent side-channel leakage.

**DSA TOY DEMO FINAL SUMMARY OUTPUT**

The source code written in Mathematica 12 and full output can be downloaded from https://1drv.ms/f/c/32525337ca8f82b9/EgW4CnRWjUhLuDrokCd5rkABQwE83kSpQC-teRb2-OD8qQ?e=aWIehz

=== **FINAL SUMMARY** ===

**Key Generation**: ✓ COMPLETED

**Signing**: ✓ COMPLETED (message: " Hello PQC! with FO & IR ")

**Verification (original)**: ✓ ACCEPTED

Verification (tampered): ✗ REJECTED*

Implicit Rejection: ✓ WORKING

All three sections were successfully completed by the participants in this study.

=== **DS DEMO COMPLETE** ===

Elapsed computing time (seconds) = 0.2031117